\documentstyle[ fleqn,epsfig]{article}

\begin{document}
\title{
\vspace{-8mm}
\rightline{\small HUB--EP--97/84}
\vspace{-2mm}
Modelling the QCD Phase Transition with an Effective Lagrangian
       of Light and Massive Hadrons
      }
\author{N.~O.~Agasyan$^1$
\thanks{Supported by 
Deutsche Forschungsgemeinschaft under grant 436 RUS 113/29}\, , D.~Ebert$^2$
\thanks{Supported in part by the Heisenberg-Landau Program}\, 
and
E.--M.~Ilgenfritz$^2$  
\thanks{Supported by 
Deutsche Forschungsgemeinschaft under grant Mu932/1-4}
      \\ \\
{$^1$ \it Institute of Theoretical and Experimental Physics,}\\{\it B. 
Cheremushkinskaya 25, 117218 Moscow, Russia}  \\ \\
{$^2$ \it Institut f\"ur Physik, Humboldt-Universit\"at zu Berlin,}\\
{\it Invalidenstrasse 110, D-10115, Berlin, Germany}
       }

\maketitle
\begin{abstract}
The temperature dependence of quark and gluon condensates in QCD 
as precursor of the chiral and deconfining phase transition is
modelled with a conformally extended non--linear $\sigma$--model
including broken chiral and scale invariance. The model is further
enlarged by including (free) heavier  hadrons. Within this frame we
then study the
interplay of QCD scale breaking effects and heavier hadrons in chiral
symmetry restoration.  

\end{abstract}

\section{Introduction}
Quantum Chromodynamics (QCD) with light quarks possesses, 
at the classical level,
the remarkable feature of chiral and conformal symmetry.
At low energy, these symmetries are broken by the non--perturbative
vacuum structure which is signalled by the appearance of nonvanishing
quark and gluon condensates, $\langle 0 | \overline{q}q |0 \rangle$ and
$\langle 0 | G_{\mu\nu}^a G_{\mu\nu}^a | 0 \rangle$, respectively \cite{SVZ}.
As a result of spontaneous breakdown of chiral symmetry there
arise light Goldstone bosons (pions) which are the relevant hadronic 
degrees of freedom of low--energy QCD. On the other hand, 
at higher temperature
the non--perturbative vacuum structure partially disappears which
manifests itself in a phase transition from hadronic matter with 
confined quarks and gluons to an interacting 
plasma of deconfined quarks and gluons. 
In particular, it is important to investigate the temperature behaviour
of the quark condensate in order to determine the critical temperature $T_c$
at which the chiral symmetry becomes restored. This question has been
extensively studied in lattice QCD \cite{DeTar} 
from first principles.
A natural question
in the context of particular models
is what influence the existence of heavier (non--Goldstone) mesons 
$\sigma$, $\rho$, $\omega$, ... and baryons $N$, $\Delta$, ... has on the
actual chiral symmetry restoration temperature $T_c$. Finally,
taking into account also the temperature--dependent gluon condensate, it is
interesting to consider the interplay of temperature effects 
on the condensates and other thermodynamic quantities.

For the study of non--perturbative condensates we use in this paper
the effective Lagrangian approach.   Effective meson Lagrangians provide a
compact and extremely useful method to summarize low--energy theorems of QCD 
\cite{Goldberg}. 
They incorporate the broken global chiral and scale symmetries of QCD.
In particular, to mimic the QCD scale anomaly \cite{Crewther}
\begin{equation}\label{eq:first_scale_anomaly}
\langle \Theta_{\mu\mu}^{QCD} \rangle = 
\langle \frac{\beta(g)}{2 g} G_{\mu\nu}^a G_{\mu\nu}^a \rangle  \, , 
\end{equation}
one introduces a scalar chiral singlet dilaton--glueball field $\chi$
with an interaction potential $V(\chi)$ \cite{Schechter} 
so that, via Noether's theorem
$\Theta_{\mu\mu} \propto \chi^4$ and (\ref{eq:first_scale_anomaly}) is
satified for $\chi_0=\langle 0 | \chi | 0 \rangle \ne 0$. As proposed in 
\cite{Elli,Campbell} the kinetic and symmetry breaking mass terms of standard chiral 
meson Lagrangians have then to be multiplied by suitable powers of 
$\chi/\chi_0$ in order to reproduce the scaling 
behaviour of the analogous terms of the QCD Langrangian. 
These ideas have 
incited a number of investigations of
the chiral and scaling behaviour of 
low--energy Lagrangians at finite 
temperature 
\cite{Brown,Patkos,Kusaka,Heinz,Ebert,Ellis}.

Here, we will start with a non--linear (scaled) $O(4)$ $\sigma$--model which
is isomorphic  to a $SU(2) \times SU(2)$ chiral 
Lagrangian. The model contains one scalar ($\sigma$) and three pseudoscalar
$\overrightarrow{\pi}$ fields, constrained by the scaled condition
$\sigma^2 + \overrightarrow{\pi}^2=f_{\pi}^2 \left(\frac{\chi}{\chi_0}\right)^2$
where $f_{\pi}$ is the pion decay constant. In the path integral evaluation 
of the thermodynamic potential this constraint can be rewritten by 
introducing a Lagrange multiplier field $\lambda(x)$. This allows to perform
the remaining Gaussian integration over the pion fields {\it exactly}
and to apply
a saddle point approximation to the $\lambda$ integral.
Finally, to model the temperature effects of heavier hadrons, 
we add the free Lagrangian of these particles. 
As discussed in \cite{GerberLeutwyler} (without
studying the interplay with the gluon condensate) their influence would
lead to a lowering of $T_c$ by about $10$ percent. Therefore, a 
generalization of the model  which includes the 
effect
of the gluon condensate 
on the low--lying hadronic states (with mass, say, $\le 2$ GeV)
is worthwile. 
To couple these non--Goldstone particles to the remaining degrees of
freedom in our model,
we have to adopt a
simple recipe defining 
the scaling properties of their masses
with $\chi$. In the spirit of
Ref. \cite{GerberLeutwyler}, 
we use a simple--minded additive quark model to determine
the chiral--symmetric and non--symmetric pieces of hadron masses. This leads
to a definite scaling prescription for hadron masses
$M_h \rightarrow M_h[\chi/\chi_0]$. Notice that the finite temperature
glueball dynamics might influence the chiral symmetry restoration, too
\cite{Campbell,Heinz}. In order to clarify this point and to compare the thermal
effect of glueballs with that of the other heavy hadrons, we include
quantum fluctuations of the glueball field. The effective masses $M_{\chi}$
and $M_{eff}$ of glueballs and pions are then obtained in a self--consistent
way from saddle point equations derived for the thermodynamical potential
(free energy) of the model.

The structure of the paper is as follows. 
In sect. 2 we introduce the effective low--energy meson Lagrangian with
broken chiral and scale invariance discussing in some detail the links
between chiral- and scale--symmetry breaking on the basis of the low--energy
theorem for the gluon condensate in the presence of non--vanishing quark
masses \cite{NSVZ}. Moreover, the 
scaling behaviour of the masses,
which is different for Goldstone and
non--Goldstone particles,
is set up. In sect. 3
we calculate the free energy density of the extended 
non--linear $\sigma$--model including pions, 
heavier hadrons and glueballs and determine the 
temperature dependent quark and gluon condensates
$\langle \langle \overline{q} q \rangle \rangle _{|T} \propto \sigma(T)$
(including the effect of heavier hadrons) and
$ \langle \langle G^2 \rangle \rangle _{|T} \propto \chi^4 $
from the saddle point equations for the $\sigma$ and $\chi$ fields, 
respectively. Finally, sects. 4 and 5 contain the discussion of numerical
results and the conclusions.

\section{The effective hadron Lagrangian}

\subsection{Symmetry breaking and scale anomaly}

In order to model the approximate chiral and scale invariance of QCD 
we consider an extended $O(4)$ non--linear $\sigma$--model containing one scalar
$\sigma$- and three pseudoscalar $\overrightarrow{\pi}$-fields 
as well as a scalar isoscalar
dilaton (glueball) field $\chi$. The corresponding Lagrangian is given in
Euclidean notation by
\begin{equation}\label{eq:eff_Lagrangian}
L(\sigma,\overrightarrow{\pi},\chi) = 
\frac12 (\partial_{\mu} \sigma)^2 +
\frac12 (\partial_{\mu} 
\overrightarrow{\pi})^2 +
\frac12 (\partial_{\mu} \chi)^2 +
V(\chi) + V_{SB}(\sigma,\chi)  ~,
\end{equation}
where the fields satisfy the 
rescaled
chiral constraint
\begin{equation}\label{eq:constraint}
\sigma^2 + \overrightarrow{\pi}^2 
= ~f_{\pi}^2 ~\left(\frac{\chi}{\chi_0}\right)^2   ~,
\end{equation}
with $f_{\pi}=93$ MeV being the pion decay constant, and where
$\chi_0=\langle 0|\chi|0 \rangle$ has been introduced for dimensional
reasons. This is the actual vacuum expectation value of the field
$\chi$ which will 
slightly
depend on the amount of explicit breaking of chiral
symmetry.

In order to take into account the gluon contribution to the QCD
scale anomaly we have included the dilaton interaction potential
\cite{Schechter}
\begin{equation}\label{eq:chi_potential}
V(\chi) = ~K ~\chi^4 ~\left(~\log \left(\frac{\chi}{\chi_q}\right) 
			 - ~\frac14 ~\right)
\end{equation}
which takes its minimum value at $\chi=\chi_q$.
Finally, $V_{SB}(\sigma,\chi)$ is the scaled chiral symmetry breaking term
of scaling dimension $3$
\begin{equation}\label{eq:V_symm_breaking}
V_{SB}(\chi,\sigma)= -c ~\sigma \left(\frac{\chi}{\chi_0} \right)^2  
\end{equation}
to be added to $V(\chi)$.
It should be noted that due to the constraint (\ref{eq:constraint}) 
the field 
$\sigma = \sigma(\chi,\overrightarrow{\pi})$ is not an independent 
degree of freedom. 
For instance, omitting at tree level the pion field 
(giving rise to the pion loop contributions), we have 
$\sigma(\chi,\overrightarrow{0})=f_{\pi} \frac{\chi}{\chi_0}$. The total
potential
\begin{equation}\label{eq:total_potential}
V_{tot}(\chi) = 
V(\chi) + V_{SB}(\chi,\sigma(\chi,\overrightarrow{0}))
\end{equation}
has then its minimum shifted to 
\begin{equation}\label{eq:chi_shift}
\chi_0 = \chi_q + \frac{3}{4 ~K ~\chi_q^3} ~c ~f_{\pi} + O(c^2) .
\end{equation}
The parameters $K$, $\chi_0$ and $c$ will be specified
later by considering the vacuum energy density
(bag constant), the glueball mass and the pion mass.

From eqs.(\ref{eq:chi_potential},\ref{eq:V_symm_breaking},\ref{eq:total_potential})
we get the trace anomaly 
\begin{eqnarray}\label{eq:trace_anomaly}
\Theta_{\mu\mu}^{eff} & = & 4 ~V_{tot}(\chi) 
-   \chi ~\frac{\partial V_{tot}(\chi)}{\partial \chi} \nonumber \\
& = & - ~K ~\chi^4 -  ~c ~f_{\pi} \left(\frac{\chi}{\chi_0}\right)^3  , 
\end{eqnarray}
and the vacuum energy density
\begin{eqnarray}\label{eq:vacuum_energy}
\epsilon_{vac} & = & \frac14 \langle \Theta_{\mu\mu}^{eff}\rangle = 
			 V_{tot}(\chi_0) \nonumber \\
     & = & - \frac14 ~K ~\chi_0^4 - \frac14 ~c ~f_{\pi}  \\
     & = & - \frac14 ~K ~\chi_q^4 - ~c ~f_{\pi} + O(c^2) \, , \nonumber
\end{eqnarray}
where we have used eq. (\ref{eq:chi_shift}) in the last line. 

It is instructive to compare this with the QCD expression for $N_f=2$
\begin{equation}\label{eq:QCD_trace_anomaly}
\Theta_{\mu\mu}^{QCD} = \frac{\beta}{2 ~g} ~G_{\mu\nu}^a G_{\mu\nu}^a 
+ m \left(~\overline{u} u + \overline{d} d~\right)  ~,
\end{equation}
where $\beta$ denotes the Gell--Mann--Low function,
\begin{equation}\label{eq:beta_function}
\frac{\beta}{2 ~g} = - \frac {b ~\alpha_s}{8 ~\pi} + .... ~, \, \, \, \, \, 
b= \frac{11}{3} N_c - \frac23 N_f ~,
\end{equation}
and where $m=m_u=m_d$ are the light current quark masses.
\footnote{We do not discuss effects due to the anomalous dimension of the 
$\overline{q} q$--operator.}

Taking into account the low-energy theorem for the gluon
condensate in the presence of non--vanishing quark masses \cite{NSVZ}
\begin{equation}\label{eq:low_energy_theorem}
\langle ~\frac{\beta}{2 ~g} G_{\mu\nu}^a G_{\mu\nu}^a ~\rangle =
\langle ~\frac{\beta}{2 ~g} G_{\mu\nu}^a G_{\mu\nu}^a ~\rangle_{|m=0}
+ 3 ~m ~\langle ~\overline{u} u + \overline{d} d ~\rangle
\end{equation}
which relates 
the shift of the gluon condensate compared to
the massless case to the quark condensate,
one obtains for the vacuum energy density the form
\begin{equation}\label{eq:vacuum_energy_QCD}
\epsilon_{vac}^{QCD} = \frac14 ~\langle ~\Theta_{\mu\mu}^{QCD} ~\rangle
= \langle ~\frac{\beta}{8 ~g}  G_{\mu\nu}^a G_{\mu\nu}^a ~\rangle_{|m=0}
+ m ~\langle ~\overline{u} u + \overline{d} d ~\rangle ~. 
\end{equation}
The comparison with eqs. (\ref{eq:vacuum_energy}) and (\ref{eq:vacuum_energy_QCD})
suggests the following identifications:
\begin{equation}\label{eq:Gluon_cond}
\langle ~\frac{\beta}{2 ~g}  ~G_{\mu\nu}^a G_{\mu\nu}^a ~\rangle_{|m=0}  = 
- ~K ~\langle ~\chi^4 ~\rangle_{|c=0} = - ~K \chi_q^4  
\end{equation}
and
\begin{equation}\label{eq:GMOR}
m ~\langle ~\overline{u} u + \overline{d} d ~\rangle  =  - ~c ~f_{\pi} ~= 
~- ~M_{\pi}^2 ~f_{\pi}^2   ~.
\end{equation}
Eq. (\ref{eq:GMOR}) is just the 
well--known Gell--Mann--Oakes--Renner relation.
We have already used that, discarding pion loops,
$\langle \sigma \rangle = \sigma_0 = f_{\pi}$
due to the constraint (\ref{eq:constraint}). 
Moreover, by expanding in eq. (\ref{eq:V_symm_breaking})  
the 
$\overrightarrow{\pi}$ 
 {\it dependent} field $\sigma$
\begin{equation}\label{eq:sigma_by_squareroot}
\sigma = f_{\pi} ~\frac{\chi}{\chi_0} 
\sqrt{1 - 
\frac{\overrightarrow{\pi}^2}{f_{\pi}^2 (\frac{\chi}{\chi_0})^2}}
\end{equation}
to leading order in $\overrightarrow{\pi}^2$ one finds that the pion mass,
given by $M_{\pi}=\sqrt{c/f_{\pi}}$, scales as
\begin{equation}\label{eq:pi_mass_scaling}
M_{\pi} \rightarrow M_{\pi} \sqrt{\frac{\chi}{\chi_0}}  ~.
\end{equation}

\subsection{The different scaling behaviour of masses}

One of the main issues of the present work is to study the interplay 
of the quark and gluon condensates at finite temperature (near the
chiral phase transition) and their relation to the hadron spectrum.
In particular, we are interested in the question how the light $u$ and $d$
quark condensate and 
the gluon condensate will be affected by the 
inclusion both of additional massive 
hadrons 
$h = \rho, \omega, N, ...$ (represented by
bosonic or fermionic fields $\varphi_h$) and a (lowest) 
glueball state. Note that for the inclusion of non--Goldstone
particles we need to know also the scaling properties of the corresponding 
mass term. Using the relations \cite{DonoghueGolowich}
\begin{equation}\label{eq:matrix_elements_meson}
\langle H(k) | \Theta_{\mu\mu} | H(k) \rangle  =  2 ~M_H^2   ~,   
\end{equation}
\begin{equation}\label{eq:matrix_elements_nucleon}
\langle N({\bf p}) | \Theta_{\mu\mu} | N({\bf p}) \rangle  = 
  M_N ~\overline{u}({\bf p}) ~u({\bf p})   
\end{equation}
(for bosons $h=H$ and fermions $h=N$)
and taking eq. (\ref{eq:QCD_trace_anomaly})
into account, we see that the masses $M_h$ contain
chiral--symmetric and non--symmetric pieces. A more detailed investigation 
of this issue requires the use of low-energy theorems of QCD 
\cite{NSVZ,DonoghueLeutwyler} which is
outside the scope of this paper. Here we shall follow a 
procedure 
which determines
the hadron masses from a simple--minded
additive quark model as follows \footnote{For similarity and differences
compare with Ref. \cite{GerberLeutwyler}.}
\begin{equation}\label{eq:masses_by_constituents}
M_h ~= ~N_h^u ~\hat{m}_u ~+ ~N_h^d ~\hat{m}_d 
\end{equation}
where $N_h^{\alpha}$ is the number of valence quarks (or antiquarks) of type
$\alpha=u, \overline{u}, d$ and $\overline{d}$ with masses
$\hat{m}_{\alpha}$ inside hadron $h$.
$\hat{m}_{\alpha}$ denotes the
total quark mass of quark type $\alpha$ which decomposes as
\begin{equation}\label{eq:quark_mass_decomposition}
\hat{m}_{\alpha} ~= ~m_{\alpha} ~+ ~m_{\alpha,dyn}  
\end{equation}
into current and dynamically generated mass.
Let,
for illustration,
the dynamical part of the quark masses $m_{\alpha,dyn}$ be represented 
by a scale--invariant interaction term of the form 
\begin{equation}\label{eq:quark_mass_term}
L_{int} = - ~g ~\sigma ~\overline{q} q 
\simeq - ~g ~f_{\pi} ~\frac{\chi}{\chi_0} ~\overline{q} q 
~= ~- ~m_{dyn} ~\frac{\chi}{\chi_0} ~\overline{q} q   
\end{equation}
where use has been made of the approximate 
expression of eq. (\ref{eq:sigma_by_squareroot}) (omitting the
pion field) and of the Goldberger-Treiman relation
$m_{dyn}=g ~f_{\pi}$. From eq. (\ref{eq:quark_mass_term}) 
we deduce the scaling behaviour of the dynamical
quark mass
\begin{equation}\label{eq:dynamical_mass_scaling}
m_{dyn} ~\rightarrow ~\frac{\chi}{\chi_0} ~m_{dyn}  ~.
\end{equation}
In the case of Goldstone bosons (pions) one has the mass formula
\begin{equation}\label{eq:goldstone_scaling}
M_{\pi}^2 ~\propto ~m ~m_{dyn}  
\end{equation}
which 
formally
reproduces the 
scaling behaviour 
of eq. (\ref{eq:pi_mass_scaling}).
For the non--Goldstone particles, on the contrary, we obtain 
from eqs. 
(\ref{eq:masses_by_constituents},
\ref{eq:quark_mass_decomposition},
\ref{eq:dynamical_mass_scaling})
the scaling prescription
\begin{equation}\label{eq:non_goldstone}
M_h ~\rightarrow ~M_h[\chi/\chi_0]
\end{equation}
with some well--defined functions $M_h$ of the ratio $\chi/\chi_0$
($M_h$ should take for $\chi=\chi_0$ the known value $M_{h0}$). 
In order to include non--Goldstone hadrons we shall simply use 
a free
Lagrangian $L^{h}(\varphi_h,M_h[\chi/\chi_0])$ for each
particle type.
This corresponds to the assumption that interaction terms with pions 
lead to higher order
derivative terms in the effective Lagrangian of the pion sector. 
Such terms will be neglected in thermodynamical considerations at low
temperatures studied here. The same holds for interactions 
between non--Goldstone particles because of their low thermal density
below the critical temperature. 

\section{Chiral and conformal thermodynamics}

\subsection{The free energy density of the extended non--linear $\sigma$--model}

In the following we shall study the thermodynamical properties of
the system of pions, glueballs and free non--Goldstone hadrons 
described by the Lagrangian
\begin{equation}\label{eq:L_full}
L = L(\sigma,\overrightarrow{\pi},\chi) 
+ \sum_{h} L^{h}(\varphi_h,M_h[\chi/\chi_0])  ~.
\end{equation}
The partition function of the Lagrangian (\ref{eq:L_full})
reads
\begin{eqnarray}\label{eq:partition_fct}
Z & = & N \int_{\epsilon-i\infty}^{\epsilon+i\infty} D\lambda
  \int  D\mu(\sigma,\overrightarrow{\pi},\chi,\varphi_h) \nonumber \\
 &  & \exp \left( - \int_0^{\beta} d\tau \int d^3 {\bf x}
    \left(~\frac{1}{2}~\left(\partial_{\mu}\sigma~\right)^2  
  + \frac{1}{2}~\left(\partial_{\mu}\overrightarrow{\pi}~\right)^2 
   \right. \right.  \nonumber \\
 & & \left. \left. + \lambda ~\left(~\sigma^2 + ~\overrightarrow{\pi}^2
    -~f_{\pi}^2 ~\left(\frac{\chi}{\chi_0}\right)^2~\right) \right. \right. \\
 & & + \left. \left. \frac{1}{2}~\left(\partial_{\mu}\chi~\right)^2 
     + V_{tot}(\sigma,\chi) 
	\right. \right. \nonumber \\
 & & + \left. \left. \sum_{h} L^{h}(\varphi_h,M_h[\chi/\chi_0]) \right) \right)
 \nonumber
\end{eqnarray}
where $D\mu(...)$ denotes the integration measure of the physical fields. 
$\beta=1/T$
is the inverse temperature, and we have rewritten the 
$\delta$-function constraint 
$\delta(\sigma^2+\overrightarrow{\pi}^2-f_{\pi}^2(\frac{\chi}{\chi_0})^2)$
following from eq.(\ref{eq:constraint}) in terms of an integral over
the auxiliary field $\lambda(x)$. Periodic (antiperiodic) boundary
conditions for boson (fermion) fields as well as the appropriate
form of the particles' Lagrangians $L^{h}(\varphi_h,M_h[\chi/\chi_0])$
are understood. In the following we shall evaluate
the path integral in eq. 
(\ref{eq:partition_fct}) in the saddle point approximation
for $\lambda$ and $\sigma$ 
(putting $\lambda(x)=\lambda=const$, $\sigma(x)=\sigma=const$ and
dropping the corresponding integrations) but 
shall keep the 
quantum fluctuations of the glueball field 
$\chi(x)=\overline{\chi}~+~\tilde{\chi}(x)$ \footnote{Note that the 
saddle point approximation in $\lambda$ corresponds to the leading 
order in a $1/N$ expansion of the $O(N)$ $\sigma$--model 
\cite{Meyer-Ortmanns}.}. 
We expand the integrand up to terms $O(\tilde{\chi}^4)$. Performing 
first the Gaussian integration over the fields $\overrightarrow{\pi}$
and $\varphi_h$ we obtain ($x=(\tau,{\bf x})$) 
\begin{eqnarray}\label{eq:new_partition_fct}
Z & = &  N^{\prime} \int D\tilde{\chi} 
   ~\exp 
   \left(
   - \int_0^{\beta} d\tau \int d^3 {\bf x} 
   \left( 
   \Phi[\lambda,\sigma,\overline{\chi}]
   ~+~\frac12(\partial_{\mu}\tilde{\chi})^2  \right. \right. \nonumber \\
 & &  \left. \left. 
   ~+~ \tilde{\chi}
	       ~\frac{\partial \Phi}{\partial \overline{\chi}}
   ~+~\frac12~\tilde{\chi}^2 
	       ~\frac{\partial^2 \Phi}{\partial \overline{\chi}^2} 
   ~+~\frac{1}{3!}~\tilde{\chi}^3
	       ~\frac{\partial^3 \Phi}{\partial \overline{\chi}^3}
   ~+~\frac{1}{4\mathrm{!}}~\tilde{\chi}^4
	       ~\frac{\partial^4 \Phi}{\partial \overline{\chi}^4}	
    \right)
    \right)
\end{eqnarray}
with
\begin{eqnarray}\label{eq:abbreviation}
\Phi[\lambda,\sigma,\overline{\chi}]  & = &
 ~\lambda 
 \left( 
 ~\sigma^2 ~-~f_{\pi}^2 ~\left(\frac{\overline{\chi}}{\chi_0}\right)^2 
 ~\right) 
 ~+~ V_{tot}(\sigma,\overline{\chi})  \nonumber \\
  + & & \frac32 ~\left(\mathrm{log} ~(-\partial^2 + 2~\lambda)\right)_{(x,x)} 
 \\
 + & & \sum_{mesons~h} g_h 
 ~\left(\mathrm{log} 
 ~(-\partial^2 + M_h[\overline{\chi}/\chi_0]^2)\right)_{(x,x)} \nonumber \\
 - & & \sum_{baryons~h} g_h 
 ~\left(\mathrm{tr_{Dirac} ~log} 
 ~(\hat{\partial} + M_h[\overline{\chi}/\chi_0])\right)_{(x,x)} \nonumber 
 \,  .
\end{eqnarray}
Here, $g_h$ are statistical 
degeneracy factors
($g_p=g_n=4, g_{\omega}=3, g_{\rho}=9$ etc.). The first logarithmic
term
in eq.(\ref{eq:abbreviation}) leads just to the partition function 
of the free relativistic Bose gas of pions with an effective 
(temperature dependent) pion mass
\begin{equation}\label{eq:m_eff_pion}
M_{eff}^2 = 2 ~\lambda   ~.
\end{equation}

In order to get rid of the term linear in $\tilde{\chi}$ in
eq.(\ref{eq:new_partition_fct}) let us choose $\overline{\chi}$ to satisfy
the following saddle point equation
\begin{equation}\label{eq:saddle_for_chi}
\frac{\partial \Phi}{\partial \overline{\chi}} = 
- 2~\lambda~f_{\pi}^2 ~\frac{\overline{\chi}}{\chi_0^2}
+\frac{\partial V_{tot}}{\partial \overline{\chi}}
\pm \frac{\partial}{\partial \overline{\chi}} 
\sum_h g_h \mathrm{tr~log} (...M_h^{(2)}[\frac{\overline{\chi}}{\chi_0}]) = 0 .
\end{equation}
The last sum refers to 
bosonic (fer\-mio\-nic) had\-rons 
and the arguments $...$ of the functional logarithms
stand for their 
respective wave operators.  
Moreover, to simplify further calculations, 
we shall take into account the interactions of massive 
hadrons with the glueball field $\tilde{\chi}$ only up to first order, 
discarding 
(on the level of first order perturbation theory)
also the term 
$O(\tilde{\chi}^3)$. In this approximation nucleons do not contribute to the
coefficient of the
$O(\tilde{\chi}^2)$ term due to their Yukawa--type coupling to the glueball
field $\tilde{\chi}$, and the massive hadron contribution in the 
coefficient of the $O(\tilde{\chi}^4)$ term can also be discarded.
On the other hand, the interaction term 
$\frac{1}{4\mathrm{!}} \tilde{\chi}^4 
\left( \partial^4
~V_{tot}(\sigma,\overline{\chi})/
\partial ~\overline{\chi}^4 \right)$ 
leads, in the mean field approach, to a dynamical contribution to the glueball
mass. To see this, it is convenient to rewrite the $O(\tilde{\chi}^4)$
term by introducing an auxiliary field $\rho$ giving rise to a Yukawa--type
coupling $\rho\tilde{\chi}^2$.   
Indeed, applying a usual Gaussian integration we have
\begin{eqnarray}\label{eq:hubbard_strato}
 & & \exp \left(-\int_0^{\beta} d\tau \int d^3 {\bf x} 
\frac{3}{4\mathrm{!}} ~V_{tot}^{(4)} ~\tilde{\chi}^4 \right) \nonumber \\
& \propto & \int D\rho \exp \left(-\int_0^{\beta} d\tau 
\int d^3 {\bf x} 
\left(-~\frac12 ~\frac{1}{V_{tot}^{(4)}} ~\rho^2 
~+~ \frac12 ~\rho ~\tilde{\chi}^2 \right) \right) 
\end{eqnarray}
where the fourth derivative of the potential is given by
\begin{equation}\label{eq:fourth_derivative}
V_{tot}^{(4)}(\overline{\chi}) = 24 ~K ~ \mathrm{log} 
\left( \frac{\overline{\chi}}{\chi_q} \right) ~+~ 44 ~K  , 
\end{equation}
and the factor $3$ in the l.h.s. of eq. (\ref{eq:hubbard_strato})
accounts for the three possible pairings to express $\tilde{\chi}^4$
as a square of $\tilde{\chi}^2$. Inserting the identity 
(\ref{eq:hubbard_strato})  into eq. (\ref{eq:new_partition_fct}), we read off
an expression for the effective glueball mass \footnote{We use here 
$M_h^2[\overline{\chi}/\chi_0] 
\approx 4 m_{dyn}^2 
\left( \overline{\chi}/\chi_0 \right)^2
\approx M_{h0}^2 \left( \overline{\chi}/\chi_0 \right)^2$
neglecting small terms $O(m)$, $O(m^2)$ depending on the current mass.}
\begin{eqnarray}\label{eq:eff_glueball_mass}
M_{\chi}^2(\rho) & = &
\frac{\partial^2}{\partial \overline{\chi}^2}
\left(
-\lambda ~f_{\pi}^2 
\left(
\frac{\overline{\chi}}{\chi_0}
\right)^2
~+~ V_{tot}(\sigma,\overline{\chi})
\right) \nonumber \\
 & & + ~\sum_h ~g_h ~\frac{2 M_h^2}{\chi_0^2} 
~\Pi(M_h^2[\overline{\chi}/\chi_0])
~+~ \rho  \nonumber \\
& = & 12 ~K ~\overline{\chi}^2 ~\mathrm{log} 
\left(
\frac{\overline{\chi}}{\chi_q}
\right)
~+~ 4 ~K ~\overline{\chi}^2 ~-~ \frac{2 ~\lambda ~f_{\pi}^2}{\chi_0^2}
~-~ \frac{2 ~c ~\sigma}{\chi_0^2} \\ 
 & & + \sum_h ~g_h ~\frac{2 M_h^2}{\chi_0^2} 
~\Pi(M_h^2[\overline{\chi}/\chi_0])
~+ ~\rho \, \, .  \nonumber
\end{eqnarray}
Here, $\Pi(M^2)$ is the expression for a closed meson loop given by the sum
of $T\ne0$ and $T=0$ contributions
\begin{eqnarray}\label{eq:Pi_for_meson_loops}
\Pi(M^2) & = & \Pi_T(M^2) ~+~ \Pi_0(M^2) \, \, , \nonumber \\
\Pi_T(M^2) & = & \int \frac{d^3 {\bf p}}{(2 \pi)^2}
\frac{1}{E({\bf p})} 
\frac{1}{\exp(\beta E({\bf p})) -1} \, \, , \\
\Pi_0(M^2) & =  & \int \frac{d^3 {\bf p}}{(2 \pi)^2} 
\frac{1}{2 E({\bf p})} \, \, , \nonumber
\end{eqnarray}
with $E({\bf p})=({\bf p}^2 + M^2)^{\frac12}$ .
Inserting eq.(\ref{eq:hubbard_strato}) into eq.(\ref{eq:new_partition_fct})
we can perform the integration over the glueball field $\tilde{\chi}$.
Thus, there remains a functional integral over the $\rho$ field,
\begin{equation}\label{eq:rho_integral}
\int D\rho ~\exp 
\left( 
-\int_0^{\beta} d\tau \int d^3 {\bf x}
\left( 
-\frac{\rho^2}{2 ~V_{tot}^{(4)}} 
+ \frac12 
\left(
\mathrm{log}(-\partial^2 + M_{\chi}^2(\rho))
\right)_{(x,x)}
\right)
\right)
\end{equation}
which will be evaluated at the saddle point solving the gap equation
\begin{equation}\label{eq:rho_saddle}
\rho_0 ~=~ \frac{V_{tot}^{(4)}}{2} ~\Pi(M_{\chi}^2(\rho_0)) \, .
\end{equation}
Finally, we have 
\begin{equation}\label{eq:Z_by_F_eff}
Z = \exp \left( - \beta V_3 F_{eff}(\sigma,\lambda,\overline{\chi})
\right)
\end{equation}
where $V_3$ is the $3D$ volume and the free energy density $F_{eff}$ is
given by
\begin{eqnarray}\label{eq:F_eff}
F_{eff}(\sigma,\lambda,\overline{\chi})
 & = & V_{tot}(\sigma,\overline{\chi})
~+~ \lambda
\left(
\sigma^2 
- f_{\pi}^2\left(\frac{\overline{\chi}}{\chi_0}\right)^2
\right) ~+~ (continued) \nonumber \\
& & ~+~ F^{\pi}(M_{eff}^2) 
 ~+~ F^{\chi}(M_{\chi}^2(\rho_0)) \\
& & ~-~ \frac18 ~V_{tot}^{(4)} ~\left( \Pi(M_{\chi}^2(\rho_0)) \right)^2  
 ~+~ \sum_h ~F_h(M_h[\overline{\chi}/\chi_0]) \, \, . \nonumber
\end{eqnarray}
Here $F^{\pi}$, $F^{\chi}$ and $F^h$ are the expressions for the thermal
determinants including the zero point energy \footnote{Note that 
$\Pi(M^2)=2 \frac{\partial F}{\partial (M^2)}$.} 
\begin{equation}\label{eq:for_all}
F^{\pi,\chi,h} = F_T^{\pi,\chi,h} + F_0^{\pi,\chi,h}
\end{equation}
with
\begin{eqnarray}
F_T^{\pi}(M_{eff}^2)   & = & 3 T \int \frac{d^3 {\bf p}}{(2 \pi)^2}
\mathrm{log} \left(1 - e^{-\sqrt{{\bf p}^2+ M_{eff}^2}/T}\right) ,
\label{eq:for_pi} \\
F_T^{\chi}(M_{\chi}^2) & = & T \int \frac{d^3 {\bf p}}{(2 \pi)^2} 
\mathrm{log} \left(1 - e^{-\sqrt{{\bf p}^2+M_{\chi}^2}/T}\right) , 
\label{eq:for_chi} \\
F_T^h(M_h)             & = & \pm g_h 
			       T \int \frac{d^3 {\bf p}}{(2 \pi)^2} 
\mathrm{log} \left( 1 \mp 
	      e^{-\sqrt{{\bf p}^2   +M_h^2}/T}
	      \right)   \nonumber \\
 & \approx  & - g_h \frac{T^{\frac52} M_h^{\frac32}}{2 \pi} e^{-M_h/T}
   \,  \label{eq:for_hadrons}.
\end{eqnarray}
The upper (lower) signs in eq.(\ref{eq:for_hadrons}) correspond
to bosons (fermions) with the masses $M_h$ substituted by 
$M_h[\overline{\chi}/\chi_0]$ depending on $\overline{\chi}$.
Finally, $F_0^{\pi}$, $F_0^{\chi}$ and $F_0^h$ denote the zero
temperature contributions to the free energy 
density of the pion, glueball and hadron fields,
\begin{eqnarray}
F_0^{\pi}(M_{eff}^2) & = & 3 \int^{\Lambda_{\chi}}
\frac{d^3 {\bf p}}{(2 \pi)^2}
\frac12 \sqrt{{\bf p}^2+M_{eff}^2} , 
\label{eq:for_pions_0} \\
F_0^{\chi}(M_{\chi^2}) & = & \int^{\Lambda_{\chi}} 
\frac{d^3 {\bf p}}{(2 \pi)^2} 
\frac12 \sqrt{{\bf p}^2+M_{\chi}^2} , 
\label{eq:for_chi_0} \\
F_0^h(M_h) & = & \pm g_h \int^{\Lambda_{\chi}}
\frac{d^3 {\bf p}}{(2 \pi)^2} 
\frac12 \sqrt{{\bf p}^2+M_h^2} . \label{eq:for_hadrons_0}
\end{eqnarray}
Also here the masses $M_h$ are understood depending on 
$\overline{\chi}/\chi_0$.
Note that, due to the necessary 
momentum cut--off, 
a naive regularization (with fixed cut--off) 
would violate the scaling properties of our effective 
theory which is constructed to substitute QCD. 
In order to keep the wanted scaling behaviour, we have introduced 
a {\it rescaled} 
cut--off $\Lambda_{\chi}=\frac{\tilde{\chi}}{\chi_0} \Lambda$ 
following the analogous argumentation in the case of the Nambu--Jona--Lasinio
model \cite{Kusaka,Ripka}. Moreover, to avoid additional parameters, 
we specify the  
cut--off 
in the spirit of effective chiral theories 
by $\Lambda = 4 ~\pi ~f_{\pi} = 1.2$ GeV {\it for all fields}.

Following Ref.\cite{Meyer-Ortmanns}, we 
now adopt a suitable subtraction
procedure which guarantees the following two renormalization 
conditions at $T=0$,
\begin{eqnarray}\label{eq:two_renormal_cond}
M_{eff}^2(\lambda_0) & = & 2 ~\lambda_0 
~= ~M_{\pi}^2 ~ = ~\frac{c}{f_{\pi}}  ~,  \\
\langle ~\sigma ~\rangle & = & \sigma_0 ~= ~f_{\pi}     ~.
\end{eqnarray} 
Here $\lambda_0$ 
and $\sigma_0$ are  
the known values of $\lambda$  and $\sigma$
which should extremize 
$F_{eff}$ at $T=0$.
In addition, here we have to fulfill a third renormalization condition
\begin{equation}\label{eq:third_renormal_cond}
\langle ~\chi ~\rangle = \chi_0 ~. 
\end{equation}
For this purpose let us define a {\it subtracted} expression
of the free energy density $F_{sub}(\sigma,\lambda,\overline{\chi})$
which is obtained from $F_{eff}(\sigma,\lambda,\overline{\chi})$ 
(\ref{eq:F_eff}) by replacing in the determinantal terms 
$F^{\pi} \rightarrow F_{sub}^{\pi}$ with
\begin{eqnarray}\label{eq:subtract_example}
F_{sub}^{\pi}(M_{eff}^2)  & = & F_T^{\pi}(M_{eff}^2) + F_0^{\pi}(M_{eff}^2) 
       \nonumber \\
 & & - F_0^{\pi}(M_{\pi}^2) 
 - \left( M_{eff}^2-M_{\pi}^2 \right) 
\frac{\partial F_0^{\pi}}{\partial (M_{eff}^2)}_{|T=0}  ,
\end{eqnarray}
and using analogous subtraction prescriptions for $F^{\chi}$, $F^h$ and
for the expressions $\Pi = 2 \frac{\partial F}{\partial (M^2)}$ 
\footnote{Strictly speaking, one has also to introduce subtraction 
terms 
$-(\Lambda_{\chi} - \Lambda) 
\frac{\partial F_0}{\partial \Lambda_{\chi}}_{|M_0}$ 
(with zero temperature masses $M_0$). Due to the weak dependence of 
$\Lambda_{\chi}$ on $T$ these terms turn out to be numerically very small.
We neglect them putting $\Lambda_{\chi} \rightarrow \Lambda$.} .
The subscript $T=0$ means that the derivatives have to be evaluated 
at the physical zero temperature masses, i.e. at $M_{eff}=M_{\pi}=139$ MeV
for the pions,
for the 
non--Goldstone hadrons at $M_h=M_{h0}$ as known from the 
Rosenfeld table, and at $M_{\chi}=M_{gb}$ 
being the mass of the lowest glueball state. 
Note that in eqs.(\ref{eq:eff_glueball_mass}) and (\ref{eq:rho_saddle})
$\Pi(M^2)$ has to be replaced by the subtracted expressions $\Pi_{sub}(M^2)$.

Then, the thermal averages, abbreviated as
$\sigma(T)= \langle \langle ~\sigma ~\rangle \rangle_T$,
$\lambda(T)= \langle \langle ~\lambda ~\rangle \rangle_T$ and
$\chi(T)= \langle \langle ~\chi ~\rangle \rangle_T$,
are obtained  from solving the 
following saddle point equations
\begin{equation}\label{eq:gap_equations}
 \frac{\partial F_{sub}}{\partial \sigma} = 0 ~~~, \,  
~\frac{\partial F_{sub}}{\partial \lambda} = 0 ~~~, \,
~\frac{\partial F_{sub}}{\partial \overline{\chi}} = 0 ~~~,  
\end{equation}
taking the derivatives at 
$\sigma=\sigma(T)$, $\lambda=\lambda(T)$ and 
$\overline{\chi}=\chi(T)$. Note that in eq.(\ref{eq:gap_equations})
we have replaced the ''classical'' value $\overline{\chi}$ by a
function $\chi(T)$ which satisfies a saddle point equation including now the
contribution of the glueball loop.
\footnote{For a justification of this
procedure in terms of quantum (loop) corrections to the effective action
we refer to Refs. \cite{Iliopoulos},\cite{Heinz}.}

The first two equations take a simple form
\begin{equation}\label{eq:sigma_gap_first}
2 ~\lambda(T) ~\sigma(T) =  c 
~\left(
~\left(
\frac{\chi(T)}{\chi_0}
\right)^2  
+ \frac{2}{\chi_0^2} 
\frac{\partial F_{sub}^{\chi}}{\partial (M_{\chi}^2)}\right)
\end{equation}
and
\begin{equation}\label{eq:lambda_gap_second}
\sigma(T) = f_{\pi} ~\frac{\chi(T)}{\chi_0} 
~\left( 
1~- ~\frac{2}{f_{\pi}^2}
     ~\frac{\chi_0^2}{\chi(T)^2}
      ~\frac{\partial F_{sub}^{\pi}}{\partial (M_{eff}^2)} 
~+ ~\frac{2}{\chi(T)^2} 
     \frac{\partial F_{sub}^{\chi}}{\partial (M_{\chi}^2)} 
\right)^{\frac12}  .
\end{equation}
No massive (non--Goldstone) hadrons 
except the glueballs contribute here because
their mass does not depend on $\lambda$. Clearly, eq.(\ref{eq:lambda_gap_second}
) is just the thermal average of the constraint (\ref{eq:constraint}) evaluated in the saddle point approximation. Notice also that the glueball loop
contributes here with a different sign than the pion loop.

It is instructive to consider for a moment the chiral limit $c=0$. 
Then eq.(\ref{eq:sigma_gap_first}) admits
the two possibilities, 
\begin{equation}
{\mathrm case~~({\it i})}~~~ 
\lambda(T) = 0 ~~ , ~~\sigma(T) \neq 0 
\end{equation}
and
\begin{equation}
{\mathrm case~~({\it ii})}~~~
\lambda(T) \neq 0 ~~ , ~~\sigma(T) = 0  ~.
\end{equation}
Clearly, case ({\it i}) describes spontaneous breakdown of chiral symmetry
with a nonvanishing order parameter $\sigma(T)$.
One expects that this holds for 
temperatures $T < T_c$, where $T_c$ is the
critical temperature, with vanishing $\lambda(T)$ 
reflecting the existence of a 
massless Goldstone pion. On the contrary, case ({\it ii}) 
should hold for $T > T_c$
where chiral symmetry is restored and pions become massive. 
The situation
should be qualitatively similar also for $c \neq 0$. 

Inserting the solutions $\sigma(T)$, $\lambda(T)$ and $\chi(T)$ into 
$F_{sub}$ yields the actual free energy density as a function
of temperature
\begin{equation}\label{eq:FofT}
F(T) = F_{sub}(\sigma(T),\lambda(T),\chi(T))  ~.
\end{equation}
By definition, the subtracted quantity $F(T=0)$ is
just the vacuum energy density 
(\ref{eq:vacuum_energy}), 
associated to the gluon and quark
condensates through eq.(\ref{eq:vacuum_energy_QCD}). Indeed, using 
$\chi(T=0)=\chi_0$, 
$\lambda(T=0)=\lambda_0=\sqrt{M_{\pi}^2/2}$,
$\sigma(T=0)=f_{\pi}$ and 
$c=f_{\pi} M_{\pi}^2$, we obtain
\begin{eqnarray}\label{eq:Fof0}
F(T=0) & = & V_{tot}(f_{\pi},\chi_0) \nonumber \\
 & = & - \frac14 ~K ~\chi_0^4 - \frac14 M_{\pi}^2 ~f_{\pi}^2   \\
 & \approx  &  - \frac14 ~K ~\chi_q^4 - M_{\pi}^2 ~f_{\pi}^2 + O(M_{\pi}^4) 
 ~~~  .  \nonumber
\end{eqnarray}
We will now determine the parameters $K$ and $\chi_0$ ($\chi_q$) in
eq.(\ref{eq:Fof0}) taking the bag constant $B$ and some lowest glueball 
mass $M_{gb}$ at $T=0$
as two reference scales \cite{Heinz}
\begin{equation}\label{eq:bagconst}
B = - ~\epsilon_{vac} ~= ~\frac14 ~K ~\chi_0^4 + \frac14 M_{\pi}^2 ~f_{\pi}^2
\end{equation}
and
\begin{eqnarray}\label{eq:glueballmass}
M_{gb}^2 & = & ~\frac{\partial^2~V_{tot}(f_{\pi},\overline{\chi})} 
                     {\partial~\overline{\chi}^2}_{|\chi_0} 
~- ~\frac{M_{\pi}^2 ~f_{\pi}^2}{\chi_0^2} \\
 & = & ~4 ~K ~\chi_0^2 ~+ 6 ~\frac{M_{\pi}^2 ~f_{\pi}^2}{\chi_0^2} ~. 
 \nonumber
\end{eqnarray}

For the bag constant we choose
$B^{\frac14} = 240$ MeV, which is compatible with the gluon 
condensate at $T=0$ extracted from QCD sum rules \cite{SVZ}.
For the glueball mass we take  $M_{\chi} = 1.6$ GeV, a value
motivated by lattice results \cite{Bitar}. 
Experimental searches favour candidates in the mass 
region of $1.5 ... 1.8$ GeV \cite{RosenfeldTable}. 

\subsection{Quark and gluon condensates at $T \neq 0$}

Let us next
investigate the influence
of the glueball field and of the heavier (non-Goldstone) 
nonstrange
hadrons on the chiral condensate of $u$ and $d$ quarks. 
Primarily, the chiral condensate is obtained as the logarithmic derivative 
of the partition function with respect to the
current quark mass $m$,
\begin{equation}\label{eq:dFdm}
\langle \langle ~\overline{q} q ~\rangle \rangle_T = - \frac{T}{V} 
\frac{\partial \log Z}{\partial m} = \frac{\partial F(T)}{\partial m}  ~.
\end{equation}
With eq.(\ref{eq:GMOR}) we have 
\begin{equation}\label{eq:mpi_instead_of_m}
\frac{\partial}{\partial m}= 
- \frac{\langle ~\overline{u} u ~+~ \overline{d} d ~\rangle_0}{f_{\pi}}
\frac{\partial}{\partial~c} =
- \frac{\langle ~\overline{u} u ~+~ \overline{d} d ~\rangle_0}{f_{\pi}^2}
\frac{\partial}{\partial~M_{\pi}^2} ~~, 
\end{equation}
such that 
\begin{equation}\label{eq:ratio_first}
\frac{\langle \langle ~\overline{q} q ~\rangle \rangle_T}
     {\langle ~\overline{q} q ~\rangle_0}
= - \frac{1}{f_{\pi}} \frac{\partial~F(T)}{\partial~c}  ~,
\end{equation}
where use has been made of the relation
\begin{equation}
\frac{\partial~F(T=0)}{\partial~c}=-f_{\pi}  ~~.
\end{equation}

Calculating $\langle \langle ~\overline{q} q ~\rangle \rangle_T$ 
we have to take into account
also the $c$--dependence of the glueball mass $M_{\chi}^2$ 
(\ref{eq:eff_glueball_mass}) and of the heavy hadron masses in 
eq.(\ref{eq:non_goldstone}) via the 
current quark masses 
(\ref{eq:masses_by_constituents}) and
(\ref{eq:quark_mass_decomposition}).
From eqs.(\ref{eq:ratio_first}) and 
(\ref{eq:lambda_gap_second}) we obtain
\begin{equation}\label{eq:ratio_second}
\frac{\langle \langle ~\overline{q} q ~\rangle \rangle_T}
     {\langle ~\overline{q} q ~\rangle_0}
= \frac{\sigma(T)}{f_{\pi}} ~\left( \frac{\chi(T)}{\chi_0} \right)^2
~ +~ \Delta ~~~, 
\end{equation}
where $\Delta$ contains contributions from the glueball and the 
other massive hadron states
\begin{equation}\label{eq:Delta}
\Delta  =  \frac{2}{f_{\pi}}~\frac{\sigma(T)}{\chi_0^2} 
	~\frac{\partial~F_{sub}^{\chi}}{\partial~(M_{\chi}^2)}
   ~-~  \frac{1}{f_{\pi}} ~\sum_h 
   ~\frac{\partial~M_h^2}{\partial~c}
	~\frac{\partial~F_{sub}^h(M_h)}
			     {\partial~(M_h^2)} . 
\end{equation}
with $M_h=M_h[\chi(T)/\chi_0]$.
Here we have used 
\begin{equation}\label{eq:dMgbdc}
 \frac{\partial~(M_{\chi}^2-\rho_0)}{\partial~c}
 = - ~\frac{2~\sigma}{\chi_0^2} \, \, ,
\end{equation}
and 
\begin{eqnarray}\label{eq:dMhdc}
 \frac{\partial~M_h^2[\chi(T)/\chi_0]}{\partial~c}
  & = & 2~M_h[\chi(T)/\chi_0] 
 \frac{\partial~M_h[\chi(T)/\chi_0]}
				   {\partial~c} \nonumber \\
 & = &  -~(N_h^u+N_h^d)~\frac{2~M_h~f_{\pi}}
			{\langle ~\overline{u} u ~+~ \overline{d} d ~\rangle_0}
\left( \frac{\chi(T)}{\chi_0} \right) \, \, ,
\end{eqnarray}
where $N_h^{u}$ and $N_h^{d}$ are the numbers of light valence $u$, $d$ quarks
(or antiquarks) in the hadron $h$. 
We postpone the discussion of the last term $\Delta$ in eq.(\ref{eq:ratio_second})
and consider 
the first one at first. 
Using  eq.(\ref{eq:lambda_gap_second}) and
discarding  there the glueball loop term in the bracket as well we 
find
\begin{eqnarray}\label{eq:ratio_third}
\frac{\langle \langle ~\overline{q} q ~\rangle \rangle_T}
     {\langle ~\overline{q} q ~\rangle_0}
& = & \left(~\frac{\chi(T)}{\chi_0}~\right)^3 
~\sqrt{~1~ - ~\frac{R}{f_{\pi}^2}}
                 \nonumber \\
& = & \left(~\frac{\chi(T)}{\chi_0}~\right)^3 
~\left(~1~ - ~\frac{R}{2~f_{\pi}^2}
	      ~-~ \frac{R^2}{8~f_{\pi}^4} ~-~.....~\right) 
\end{eqnarray}
with
\begin{equation}\label{eq:R_expression}
R  =  \frac{\chi_0^2}{\chi(T)^2} 
~\frac{\partial~F_{sub}^{\pi}}{\partial~\lambda} 
_{|\lambda=\lambda(T)} \, \, .  
\end{equation}
For the special case $c=0$, when $M_{eff}(T)=M_{\pi}=0$ (putting 
$\Lambda_{\chi}=\Lambda$), we get
\begin{equation}\label{eq:special_case}
R = \frac{\chi_0^2}{\chi(T)^2} ~\frac{3~T^2}{12}  \, \, .
\end{equation}
Notice that if we set $\chi(T)/\chi_0 \rightarrow 1$  
we recover the structure of results obtained 
in chiral perturbation theory \cite{GerberLeutwyler} with inclusion of 
one-- and two--loop contributions ~\footnote{The
two--loop contribution $\propto R^2/f_{\pi}^4$ in
eq.(\ref{eq:ratio_third})
differs from the corresponding term in the $SU(2) \times SU(2)$ matrix
formulation numerically. This deviation is due to next--leading 
order terms in the $1/N$ expansion not considered here \cite{Bochkarev}.}. 
In the case $c \neq 0$ our loop expressions
contain a thermal pion mass $M_{eff}(T)=2 ~\lambda(T)$ which has to be
determined numerically in a selfconsistent way from 
eqs.(\ref{eq:gap_equations},
 \ref{eq:sigma_gap_first},
 \ref{eq:lambda_gap_second}).

Let us 
consider next at which critical temperature $T_c$ the order parameter
$\langle \langle ~\overline{q} q ~\rangle \rangle_T$ 
vanishes such that chiral symmetry is 
restored. From eq.(\ref{eq:ratio_third}) we obtain, 
assuming
$\chi(T_c) \neq 0$, that $T_c$ is determined by
\begin{equation}
R(T_c) = f_{\pi}^2  ~.
\end{equation}
Moreover, in the simplified case of massless quarks ($c=0$, $m=0$), we can use 
eq.(\ref{eq:special_case})
to derive the estimate ($\chi_0=\chi_q$)
\begin{equation}\label{eq:Tc_estimate}
T_c = 2 ~f_{\pi} ~\frac{\chi(T_c)}{\chi_0} = 186 ~\frac{\chi(T_c)}{\chi_0}
~~{\mathrm MeV} \, \, .
\end{equation}
The factor $\chi(T_c)/\chi_0$ can be expressed by the gluon condensate.
Indeed, taking the thermal average of 
eqs.(\ref{eq:trace_anomaly},\ref{eq:QCD_trace_anomaly})
we obtain for
$m = 0$, $c = 0$
\begin{equation}\label{eq:G2_condensate}
- K ~\langle \langle ~\chi^4 ~\rangle \rangle_T 
   = \langle \langle ~\frac{\beta}{2 ~g} ~G_{\mu\nu}^a ~G_{\mu\nu}^a 
    ~\rangle \rangle_T   ~,
\end{equation}
or
\begin{equation}\label{eq:G2_condensate_ratio}
\frac{\langle \langle ~G^2 ~\rangle \rangle_T}{\langle ~G^2 ~\rangle_0} = 
\frac{\langle \langle ~\chi^4 ~\rangle \rangle_T}{\chi_0^4} 
\approx \left( \frac{\chi(T)}{\chi_0}\right)^4   ~,
\end{equation}
where $G^2 = G_{\mu\nu}^a G_{\mu\nu}^a$ and the 
approximation 
       $\langle \langle ~\chi^4 ~\rangle \rangle_T 
\approx \langle \langle ~\chi ~\rangle \rangle_T^4
= \chi(T)^4$ has been 
made.
From these considerations we expect
\begin{equation}
T_c = \left(\frac{\langle \langle ~G^2 ~\rangle \rangle_T}
		 {\langle ~G^2 ~\rangle_0}
\right)^{\frac14} ~186 ~~{\mathrm MeV}  ~.
\end{equation}
As we shall find below, 
$\langle \langle ~G^2 ~\rangle \rangle_T/\langle ~G^2 ~\rangle_0 \leq 1$ 
in the region
where $\langle \langle ~\overline{q} q ~\rangle \rangle_T \neq 0$.
The gluon condensate depends only very weakly on $T$ in this region.
Therefore, the estimate $2 ~f_{\pi}$ for the chiral transition temperature
remains practically unchanged by it. 
Let us next consider the positive glueball term 
in eq.(\ref{eq:lambda_gap_second}). 
Although it is small in 
comparision with the pion loop, 
its contribution is expected to be non-negligible 
in the region $T\sim 2 ~f_{\pi}$ where the first two terms in eq.(\ref{eq:lambda_gap_second}) approximately cancel, shifting thus $T_c$ upwards.
On the other hand, as was argued in Ref. \cite{GerberLeutwyler} (without 
consideration of gluon condensation and glueballs),
the inclusion of
massive hadron states alone 
leads to a negative contribution to the
quark condensate and 
gives rise to
a lowering of $T_c$ by about $10$ percent.
In order to estimate the strength of this effect in our model we have to 
consider the additional terms in eq.(\ref{eq:ratio_second}) omitted until now.
Notice first that the arising glueball contribution in eq.(\ref{eq:Delta}) 
is - in addition to the exponential suppression with the high mass -
suppressed by the small factor $\sigma(T)$ 
compared to
the contribution of
other 
heavier hadrons and can thus be safely neglected. 
Clearly, the remaining negative hadron contributions will 
have a tendency of lowering $T_c$ and thus counteract its 
increase originating from the positive glueball 
term in eq.(\ref{eq:lambda_gap_second}). 
The numerical investigation of the interplay of 
both effects will be done in Section 4.

Finally, the temperature dependence of the gluon condensate 
(\ref{eq:G2_condensate_ratio})
is numerically determined from the third saddle point equation
$\partial F_{sub}/\partial \overline{\chi}_{|\chi(T)}=0$
in eq.(\ref{eq:gap_equations})
together with 
eqs. (\ref{eq:sigma_gap_first}) and (\ref{eq:lambda_gap_second}).

\section{Discussion of numerical results}

\begin{figure}[!htb]
\centering
\epsfig{file=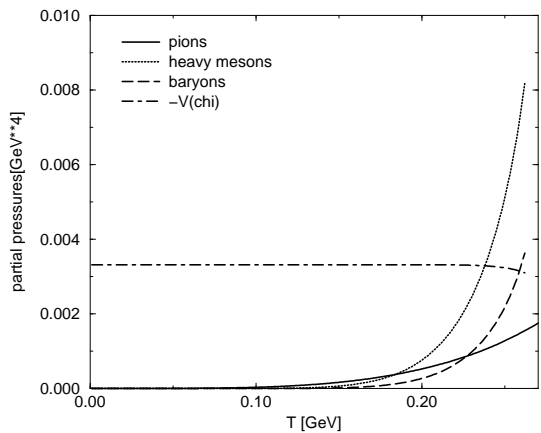,width=8.0cm,angle=0}
\caption{Temperature dependence of the partial pressures $p_i(T)$
for pions, for heavier mesons, for baryons and the contribution from the
dilaton potential $V(\chi)$.}
\label{fig:pressures}
\end{figure}
In this section we present the results of the solution of the saddle point
equations. In Fig. \ref{fig:pressures} 
we show for sake of illustration, the temperature
dependence of the thermal pressure(s) $p_i(T)$ of pions,
all heavy mesons and baryons 
as well as the tree (potential) term,
for the chiral limit $m \rightarrow 0$. Since the $\sigma$ model taken
here describes only $N_f=2$ light flavours, we have included only
non--strange heavier hadrons.
\begin{figure}[!htb]
\centering
\epsfig{file=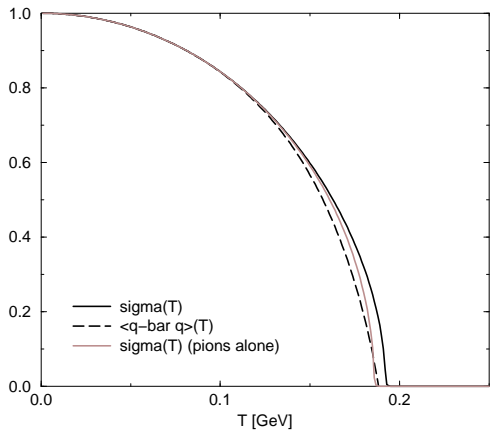,width=8.0cm,angle=0}
\caption{Temperature dependence of the thermal average $\sigma(T)$ of
our model (compared with the $\sigma$--model for pions only) 
and the quark condensate
$\langle \langle \overline{q} q \rangle \rangle_T$ for the chiral limit
using a bag constant $B^{1/4}=240$ MeV. All quantities are normalized 
to their values at $T=0$.}
\label{fig:sigma_chiral}
\end{figure}

In Fig. \ref{fig:sigma_chiral}
the thermal average $\sigma(T)$ is presented together with
the quark condensate as function of temperature in the chiral limit.
One finds that the gluon condensate
(included to mimic the 
breaking of scale invariance in QCD)
has only a negligible
influence on the behaviour of $\sigma(T)$ and on the quark condensate. 
The vanishing of $\sigma(T)$ at 
some $T_c$ is dominated by the intrinsic chiral dynamics of
pions modified by an increase of $T_c$ by about 10 MeV (6 percent) due to glueballs, but is not related to the small reduction of $\chi(T)$ with
rising
temperature. This conclusion is in accord with Ref. \cite{Heinz} 
and does 
not confirm the corresponding conjecture of Ref. \cite{Campbell}. We see that
heavier non--Goldstone hadrons indeed influence the quark condensate 
in the region of $T \le 186$ MeV due to the current quark mass dependence
of the hadron masses. They yield an effective lowering by 
about 10 MeV of the increased value of $T_c$ obtained for 
the $\sigma$--model with pions and glueballs leading to a restoration of the value $T_c \approx 2 f_{\pi}=186$ MeV obtained for pions alone.
These results are in qualitative agreement with the conclusions of Ref.
\cite{GerberLeutwyler} where pions and heavier hadrons 
have been included and contributions to the
thermodynamical potential have been calculated up to three loop level for
pions 
(while
no interplay with the gluonic degrees of freedom has been considered).
\begin{figure}[!htb]
\centering
\epsfig{file=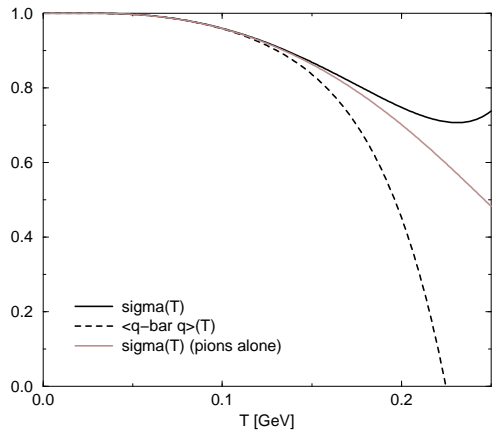,width=8.0cm,angle=0}
\caption{Same as Fig. 2
for a finite pion
mass $M_{\pi}=139$ MeV ($m=7$ MeV) using the same bag constant.}
\label{fig:sigma_massive}
\end{figure}

Fig. \ref{fig:sigma_massive} 
exhibits the same quantities for the case of finite quark masses $m=m_u=m_d\approx 7$ MeV 
(explicit chiral symmetry breaking) which are adjusted to give the pion a
mass $M_{\pi}=139$ MeV. Without QCD
scale breaking effects
taken into account
$\sigma$ is now shifted upward throughout all temperatures
and does not
play anymore the role of an order parameter.  With the coupling to the 
gluon condensate and, through the glueball mass, to the glueball loop,
$\sigma$ even bends upward for temperature $T > 220$ MeV. This can be
interpreted as 
a kind of stabilization of $\sigma$ 
by the gluonic degrees of freedom. 
A similar behaviour has been found in Ref. \cite{Ebert}.
\begin{figure}[!htb]
\centering
\epsfig{file=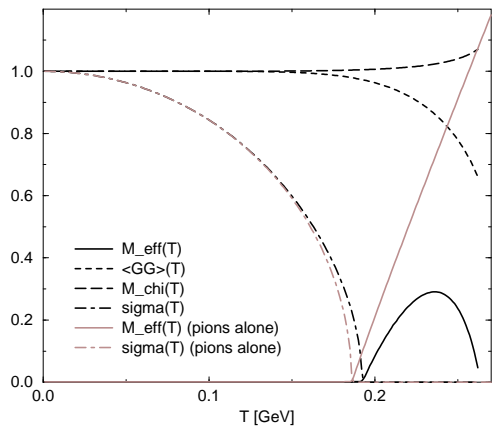,width=8.0cm,angle=0}
\caption{Temperature dependence of $\sigma(T)$, the effective pion mass
$M_{eff}(T)$, the gluon condensate 
$\langle \langle GG \rangle \rangle_T$ and the glueball mass $M_{\chi}(T)$
for the chiral limit using the same bag constant as before. The effective
pion mass is normalized to $M_{\pi}=139$ MeV, all other quantities to
their respective values at $T=0$.}
\label{fig:gluons_chiral}
\end{figure}

If we neglect the smaller glueball loop contributions we see that the
temperature dependence of $\sigma(T)$ is directly    
related to  
the ratio $\chi(T)/M_{eff}(T)$ (see eq. (\ref{eq:sigma_gap_first}). 
\begin{figure}[!htb]
\centering
\epsfig{file=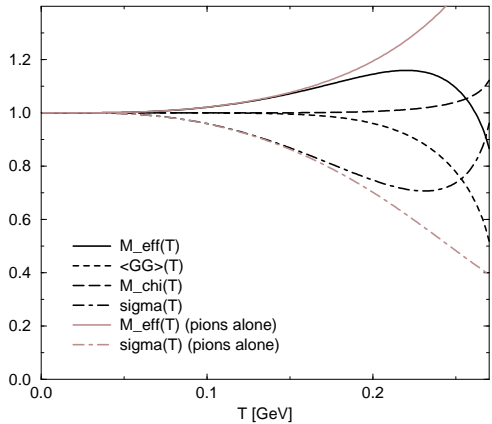,width=8.0cm,angle=0}
\caption{Same as Fig. 4 for $M_{\pi}=139$ MeV ($m=7$ MeV).}
\label{fig:gluons_massive}
\end{figure}
Finally, including the negative contributions of the other heavier
hadrons, the total quark condensate is again  strongly decreased
and vanishes at $T_c \approx 220$ MeV. In Fig. \ref{fig:gluons_chiral}
we show the
temperature dependence of the effective pion mass $M_{eff}(T)$, the
gluon condensate $ \langle \langle G^2 \rangle \rangle _{|T} \propto \chi^4 $
and the effective glueball 
mass $M_{\chi}(T)$ in the case of the chiral limit. 
All quantities are normalized to their
respective values at $T=0$, with the only exception of $M_{eff}$, which is
always normalized as $M_{eff}/M_{\pi}$, whether $c \rightarrow 0$ or not.
Notice that for $T > T_c$ there arises a non--vanishing pion mass from
the $\sigma$--model, whose 
teperature behaviour is now influenced by the gluon condensate.
The plot ends when $\lambda$ hits the constraint $\lambda \ge 0$.

Fig. \ref{fig:gluons_massive} 
presents the same quantities as Fig. 4 in the case of finite quark
and pion mass. Clearly, for temperatures $T \ge 180$ MeV the above
discussion and the figures must be taken with caution with respect to
real QCD.   From the point of view of the effective hadronic Lagrangian
one has to recall that in this temperature region higher
order derivative terms have to be included into the action and higher
loop contributions are expected to contribute.  Therefore our
results are at best 
qualitative ones in the temperature region $T > T_c$.

\section{Conclusions}

In this paper we tried to give a qualitative description of various important
aspects of temperature--dependent QCD using an effective Lagrangian approach.
Our analysis is based on a conformally extended non--linear 
$\sigma$--model which describes light pions 
coupled to the gluon condensate
and,
eventually, to
glueball thermal loop effects. 
Most importantly, this model
embodies explicit and dynamical breaking of chiral
and conformal symmetry by construction.
Heavier non--Goldstone mesons and baryons have further been
added in order to model the additional dependence of the 
condensates via heavy hadron masses.

The primary motivation was to investigate in this framework the mutual
influence of quark and gluon condensates. 
However, if the parameters are
fixed using information from zero temperature like the
vacuum energy density and the glueball mass,
we find that 
the gluon condensate is rather temperature independent in
this physical frame up to temperature $T \approx 200$ MeV. 
At somewhat higher temperature, in
the
interval up to $T=250$ MeV, it decreases to $\approx 0.75$ of the zero 
temperature value. This effect is accompanied by a (roughly inverse)
increase of the glueball mass. If the bag constant is taken smaller
($B=(140 \mathrm{ MeV})^4$ instead of $B=(240 \mathrm{MeV})^4$) 
this ''melting'' and the stronger rise of the glueball mass with temperature
set in 
at slightly lower temperature.

In the chiral limit,
the temperature dependence of the $\sigma$ field, the quark condensate
etc. is almost independent of the gluon condensate, but slightly changed by the glueball
loop contributions. It is strongly influenced by the explicit chiral symmetry
breaking through the quark mass term. Let us first summarize the 
case of the 
chiral limit
($m_u=m_d=M_{\pi}=0$). Discarding loop effects of glueballs and
heavy hadrons, the critical temperature of chiral symmetry restoration
turns out to be
$T_c=2 ~f_{\pi} \frac{\chi(T_c)}{\chi_0} \approx 186$ MeV, in view of 
$\chi(T_c)/\chi_0\approx 1$. The inclusion of the glueball loop in $\sigma(T)$ then
increases this value of $T_c$ by about 10 MeV. Next, heavy hadrons and dynamical
glueballs contribute
with different signs to the extra term $\Delta$ (cf. eq. (\ref{eq:Delta})) in the quark
condensate. Due to the large glueball mass and the small value of the factor $\sigma(T)$
near $T_c$
the respective glueball term is here numerically too small to counteract the
decrease of the quark condensate (and of $T_c$) 
caused
by heavy hadrons. In accord with  
Ref. \cite{GerberLeutwyler}, we find
a 6 percent (10 MeV) downward shift of $T_c$ which 
just compensates the corresponding increase
arising from $\sigma(T)$ in our model.
In particular, our results confirm the conclusion of Ref.
\cite{Heinz} that in the chiral limit the vanishing of the quark condensate is, above all,
dictated by the internal pion dynamics. 
This behaviour is only slightly corrected by glueballs and
the heavier 
non-Goldstone mesons and baryons and has nothing to do with a decrease 
(not to mention vanishing !) of the gluon condensate.  Thus, the
conjecture of 
Ref. \cite{Campbell} is not supported by our analysis. Note also that 
at $T > T_c$ the effective pion mass  becomes non--zero, grows to a maximum
and decreases to zero again. This behaviour is related to the beginning
decrease of the gluon condensate in the temperature interval 
$200$ MeV $ < T_c < 250$ MeV. 
However,
beyond the temperature of chiral
symmetry restoration the predictions of the 
$\sigma$--model (present here in its simplest
version) have to be taken with caution and are at best qualitative ones.

In the more realistic case of soft explicit breaking of chiral symmetry by 
finite quark and pion masses, the curve of the pion loop contributions
to the quark condensate (represented by the saddle point solution $\sigma(T)$)
is shifted upwards 
and does not clearly define a phase transition.
(This is the case even for the pure $\sigma$ model.)
Discarding the
small glueball loop contributions, this behaviour can be 
directly related to the behaviour of the ratio $\chi(T)/M_{eff}(T)$ as
can be seen from eq. (\ref{eq:sigma_gap_first}).

Nevertheless, including also non-Goldstone hadrons, their contribution
is
sufficient to
make the quark condensate 
vanish, however 
at somewhat  
higher temperature.
This happens at $T_c \approx 220$ MeV. 
Again, 
we have to emphasize
that this temperature region is beyond the region of applicability of the simple
non--linear $\sigma$--model used here 
as a corner stone of our 
extended model. From chiral perturbation 
theory \cite{GerberLeutwyler,GasserLeutwyler} it is known that higher 
order derivative  terms in the Lagrangian and higher loops cannot be neglected
at these temperatures.

Summarizing, our model exhibits in the chiral limit, for temperatures 
$T \le 200$ MeV, competing effects from glueballs (increase of $T_c$)
and heavier hadrons (decrease of $T_c$) which, because of cancellation, have however 
only a minor  influence on the chiral
symmetry restoration. The gluon condensate shows also very weak temperature
dependence in this temperature range.
On the other hand, in
the more realistic case of finite quark and pion masses, it is just the quark mass
dependence of the heavier hadron masses which 
leads to a significant lowering of the quark condensate and, finally, to its vanishing.

Moreover, noticeable variations of the gluon condensate
down to $75$ percent of the zero 
temperature value are only seen in a 
temperature region $200 \mathrm{MeV} < T_c < 250$ MeV where the simple
non--linear $\sigma$--model is not realistic anymore.   

Modelling the implicitely temperature dependent heavy hadron masses
by another dependence on the order parameters might lead to stronger and yet
more 
interesting to analyse effects for the 
detailed temperature dependence of chiral symmetry restoration. 
Such models need dynamical input from hybrid models 
including quark degrees of freedom simultaneously with the Goldstone fields.
These 
mechanisms
will be the
subject of future studies.

\section*{Acknowledgement}
One of the authors (D.E.) would like to thank M.A.Shifman, A.I.Vainshtein, M.B.Voloshin,
S.Rudaz and P.J.Ellis for useful discussions during his stay at the Institute of Theoretical
Physics of the University of Minnesota.

\end{document}